\documentclass[aps,physrev,reprint,
        superscriptaddress,
        nofootinbib, 
        amsmath,
        amssymb,]{revtex4-2}
\usepackage{graphicx}
\usepackage{xcolor}
\definecolor{blueviolet}{rgb}{0.2, 0.2, 0.6}
\definecolor{webgreen}{rgb}{0,.5,0}
\definecolor{webbrown}{rgb}{.6,0,0}
\usepackage[pdftex,
	bookmarks=false,
	colorlinks=true, 
	urlcolor=webbrown, 
	linkcolor=blueviolet, 
	citecolor=webgreen,
	pdfstartpage=1,
	pdfstartview={FitH},  
	bookmarksopen=false
	]{hyperref}

\begin{document}


\title{Unsupervised Detection of Topological Phase Transitions with a Quantum Reservoir}


\author{Li Xin}
    \email{lixinphy@bit.edu.cn}
    \affiliation{Center for Quantum Technology Research and Key Laboratory of Advanced Optoelectronic Quantum Architecture and Measurements (MOE), \\ School of Physics, Beijing Institute of Technology, Beijing 100081, China}
\author{Da Zhang}
    \affiliation{Center for Quantum Technology Research and Key Laboratory of Advanced Optoelectronic Quantum Architecture and Measurements (MOE), \\ School of Physics, Beijing Institute of Technology, Beijing 100081, China}
\author{Zhang-Qi Yin}
    \email{zqyin@bit.edu.cn}
    \affiliation{Center for Quantum Technology Research and Key Laboratory of Advanced Optoelectronic Quantum Architecture and Measurements (MOE), \\ School of Physics, Beijing Institute of Technology, Beijing 100081, China}


\date{\today}

\begin{abstract}
In quantum many-body systems, characterizing topological phase transitions typically requires complex many-body topological invariants, which are costly to compute and measure. Inspired by quantum reservoir computing, we propose an unsupervised quantum phase detection method based on a many-body localized evolution, enabling efficient identification of phase transitions in the extended SSH model. The evolved quantum states produce feature distributions under local measurements, which, after simple post-processing and dimensionality reduction, naturally cluster according to different Hamiltonian parameters. Numerical simulations show that the evolution combined with local measurements can significantly amplify distinctions between quantum states, providing an efficient means to detect topological phase transitions. Our approach requires neither complex  measurements nor full density matrix reconstruction, making it practical and feasible for noisy intermediate-scale quantum devices. 
\end{abstract}


\maketitle

\section{Introduction}
Symmetry-protected topological (SPT) phases represent a distinct class of quantum matter in many-body physics, with profound implications for both quantum computation and strongly correlated systems~\cite{Sachdev_2011, PhysRevLett.86.5188}. Unlike conventional phases characterized by spontaneous symmetry breaking, SPT phases cannot be described by any local order parameter; instead, they are defined through global properties such as entanglement entropy and many-body topological invariants (MBTI). Experimentally, their detection goes beyond standard linear-response measurements in condensed matter physics, while from a quantum information perspective it often requires full reconstruction of the system’s density matrix~\cite{PhysRevLett.118.216402, sciadvaaz3666, PhysRevB.98.035151, PhysRevResearch.2.033030}. These challenges make the identification of SPT phases—particularly in strongly correlated settings—highly nontrivial. The advent of noisy intermediate-scale quantum (NISQ) devices offers a powerful new platform for simulating quantum states and dynamics of many-body systems, thereby enabling novel approaches to studying correlated and topological phases~\cite{RevModPhys.94.015004, arute2020observationseparateddynamicscharge, Jin2025Topological, Will2025Topological, Evered2025}. In parallel, the rapid progress of machine learning provides new perspectives for identifying and classifying phase transitions in many-body physics~\cite{PhysRevB.96.144432, Wang2018KPCA, Wang_2016}. Against this backdrop, a surge of recent research has emerged in this field. Nevertheless, strongly correlated topological transitions remain challenging.

For instance, several studies have employed manifold-based approaches to identify topological quantum phase transitions. Notably, diffusion map techniques have been applied to one-dimensional topological insulators and superconductors~\cite{PhysRevB.102.134213, PhysRevLett.126.240402, RodriguezNieva2019TopoML}. However, these methods largely rely on single-particle Bloch representations of the Hamiltonian, which presuppose a well-defined Brillouin zone, and thus cannot be straightforwardly extended to strongly correlated topological systems.

Heuristic quantum algorithms on NISQ platforms have also been employed to study topological phase transitions; for instance, methods such as QCNNs have been applied to the classification of SPT phases~\cite{Cong_2019, Chen2025QEL}. These approaches offer fresh perspectives for many-body physics, yet they face clear limitations: on the one hand, they often rely on supervised learning and thus require prior knowledge of the system; on the other hand, training such quantum algorithms involves extensive parameter optimization, which remains prohibitively costly on current NISQ devices and limits their practical significance.

 A particularly important line of work builds on randomized measurement tomography~\cite{Elben2023Randomized}, also known as the classical shadow framework~\cite{Huang_2020}. These methods claim to bridge the gap between measurement data obtained from quantum devices and physically meaningful observables within a practically acceptable resource overhead, while integrating well with modern machine learning and deep neural network models, such as Transformers~\cite{yao2024shadowgptlearningsolvequantum} and diffusion models~\cite{tang2025quadim}, among other state-of-the-art generative networks. Although these approaches are, in principle, capable of addressing many-body problems—including quantum phase transitions—with polynomial complexity, the required number and precision of POVM measurements remain prohibitively demanding for current NISQ devices. As a result, most of these methods currently remain at the theoretical stage, with numerous challenges yet to be resolved for experimental implementation.

Meanwhile, recent discoveries of novel quantum states, such as dynamical phase transitions~\cite{PhysRevB.107.094307, Muniz2020DPT} and many-body localization (MBL)~\cite{Abanin_2019}, have inspired heuristic algorithms based on quantum reservoir computing (QRC)~\cite{ PhysRevLett.127.100502, PRXQuantum.3.030325, Xia_2022}, achieving notable success in classical tasks including classification and time-series prediction. However, the application of QRC to learning quantum data has not been explored. In this work, we pioneeringly introduce QRC to the study of SPT quantum states, enabling unsupervised learning of strongly correlated SPT phases and efficient, direct reconstruction of phase diagrams.

To validate our approach, we perform numerical simulations using tensor networks as well as experiments on quantum circuits. Our work can be viewed both as a comprehensive proposal tailored for NISQ devices and as a hybrid numerical framework that integrates quantum and classical algorithms.

\section{Method}

\subsection{Framework overview}

\begin{figure*}[htpb]
\centering
\includegraphics[width=1\linewidth]{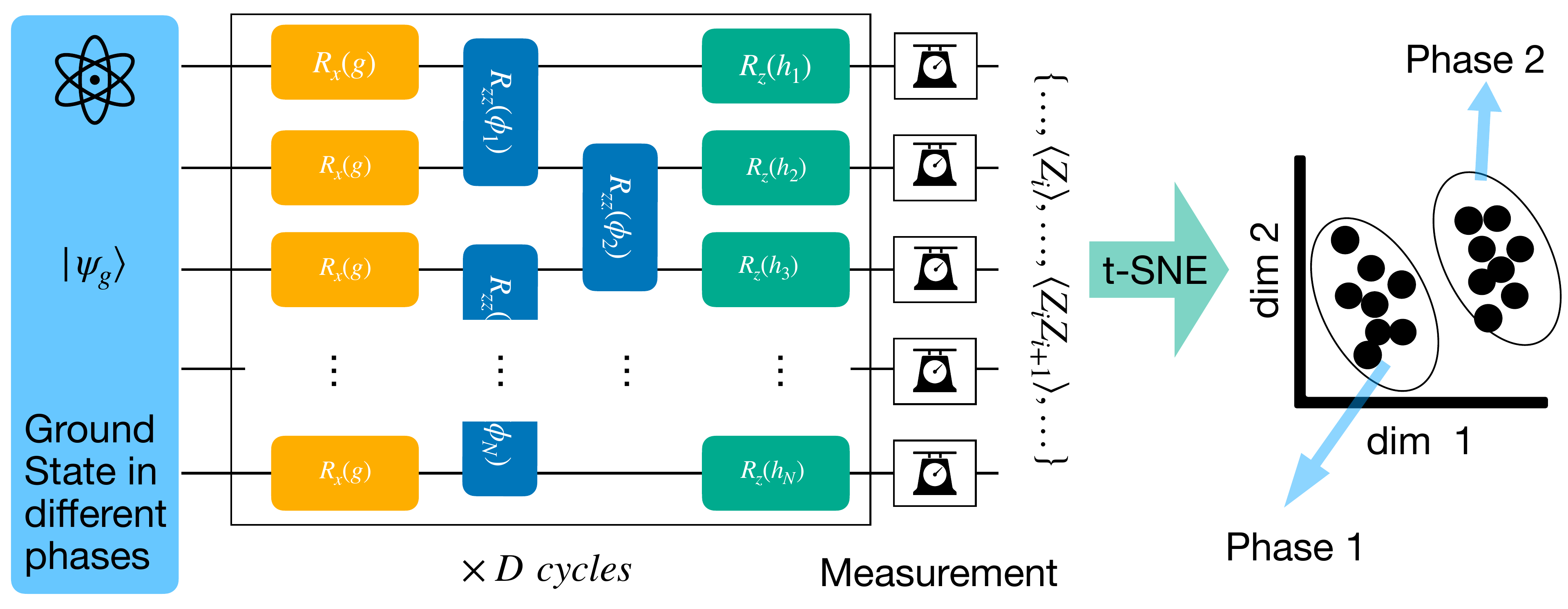}
\caption{Schematic illustration of the proposed QRC framework. We prepare a set of quantum states under different parameters, which can be obtained on NISQ devices via VQE or in analog quantum simulators by tuning system parameters. In our numerical experiments, these states are generated using DMRG and then evolved under a quantum circuit in the DTC regime. We measure only $\langle Z_i \rangle$ and $\langle Z_i Z_{i+1} \rangle$ to form feature vectors, which are subsequently visualized using t-SNE. The results show that feature vectors from different phases cluster effectively in the feature space, enabling unsupervised learning of phase transitions in strongly correlated systems.}
\label{fig:framework}
\end{figure*}

To introduce our framework, we first review QRC.

Reservoir computing (RC) is a paradigm that leverages a fixed high-dimensional dynamical system as a feature map, projecting input data into the reservoir states, which are then processed by simple linear or shallow neural networks~\cite{doi:10.1126/science.1091277, TANAKA2019100, Kobayashi2023Reservoir}. QRC generalizes this idea by using the dynamical evolution of a quantum system as the reservoir, mapping classical or quantum inputs into a high-dimensional Hilbert space, thereby exploiting quantum superposition and entanglement to enhance information processing~\cite{PhysRevApplied.8.024030}. In typical implementations, a unitary quantum circuit serves as the reservoir, with local operator expectation values, such as single-body $\langle Z \rangle$ or two-body $\langle ZZ \rangle$, measured as the reservoir output and fed into a neural network for tasks such as image recognition or time-series generation~\cite{PRXQuantum.3.030325, PhysRevLett.127.100502}. However, these tasks mostly involve classical data, leaving the latent computational potential of quantum systems largely untapped.

In this work, we propose a novel approach: quantum states from different phases are directly fed into the quantum reservoir, and local operator expectation values of the evolved states are measured for unsupervised learning, enabling efficient quantum phase detection while fully exploiting the processing power of quantum dynamics. Our reservoir is realized as a discrete-time crystal (DTC) circuit with Floquet evolution~\cite{PRXQuantum.2.030346, Mi_2021}:
\begin{equation}
U_F(g, \boldsymbol \phi, \boldsymbol h) = e^{-ig\sum_{i=1}^{N}X_i} e^{-i\sum_{i=1}^{N-1} \phi_i Z_i Z_{i+1}} e^{-i\sum_{i=1}^{N} h_i Z_i},
\end{equation}
where parameters are sampled randomly: $\phi_i \in [-1.5\pi, -0.5\pi]$, $h_i \in [-\pi, \pi]$. The X-flip parameter $g$ can be chosen from $(0, \pi)$; however, the system enters the MBL phase for $g < 0.2\pi$ or $g > 0.84\pi$, and exhibits period-doubled oscillations (discrete time crystal) for $g > 0.84\pi$. In other regions, the system thermalizes according to the eigenstate thermalization hypothesis~\cite{Deutsch_2018}.

Previous studies~\cite{zhang2025robustefficientquantumreservoir} have shown that DTC circuits can suppress thermalization, retain long-term memory, and exhibit robustness in classification tasks. We adopt DTC circuits because they scramble local information of the input states while preserving memory of the initial state, providing crucial support for unsupervised learning and enabling efficient extraction of structural features from quantum states.

Fig.~\ref{fig:framework} illustrates the schematic of our framework.

\subsection{Unsupervised Learning with t-SNE}

For a system with $L$ sites, after evolution through the quantum reservoir circuit, we measure single-site operators $\langle Z_i \rangle$ for $i=1,\dots,L$ and two-site operators $\langle Z_i Z_{i+1}\rangle$ for $i=1,\dots,L-1$. These measurements are combined into a high-dimensional vector
\begin{equation}
\mathbf{x} = (\langle Z_1 \rangle, \dots, \langle Z_L \rangle, \langle Z_1 Z_2 \rangle, \dots, \langle Z_{L-1} Z_L \rangle),
\end{equation}
and the same procedure is applied to a batch of quantum states from different phases, yielding a batch of high-dimensional vectors. We then perform unsupervised learning on these vectors to identify clusters corresponding to different quantum phases and reconstruct the phase diagram.

We employ t-distributed stochastic neighbor embedding (t-SNE), a nonlinear dimensionality reduction technique that preserves local neighborhood structure in the low-dimensional embedding~\cite{NIPS2002_6150ccc6}. Given high-dimensional vectors $\{\mathbf{x}_1, \dots, \mathbf{x}_N\}$, t-SNE defines the conditional probability in the high-dimensional space as
\begin{equation}\label{equ:pca}
p_{j|i} = \frac{\exp\big(-\|\mathbf{x}_i - \mathbf{x}_j\|^2 / 2\sigma_i^2\big)}{\sum_{k\neq i} \exp\big(-\|\mathbf{x}_i - \mathbf{x}_k\|^2 / 2\sigma_i^2\big)},
\end{equation}
where $\sigma_i$ is determined adaptively according to a user-defined perplexity. The probabilities are then symmetrized:
\begin{equation}
p_{ij} = \frac{p_{i|j} + p_{j|i}}{2N}.
\end{equation}
In the low-dimensional embedding (typically 2D), the similarity between points $\mathbf{y}_i$ and $\mathbf{y}_j$ is modeled using a Student-t distribution with one degree of freedom:
\begin{equation}
q_{ij} = \frac{(1+\|\mathbf{y}_i - \mathbf{y}_j\|^2)^{-1}}{\sum_{k \neq l} (1+\|\mathbf{y}_k - \mathbf{y}_l\|^2)^{-1}}.
\end{equation}
The low-dimensional embedding is obtained by minimizing the Kullback-Leibler divergence between the high- and low-dimensional distributions:
\begin{equation}
\mathcal{L}_{\rm KL} = \sum_{i \neq j} p_{ij} \log \frac{p_{ij}}{q_{ij}}.
\end{equation}
After optimization, similar quantum states cluster together in the low-dimensional space, while dissimilar states are well separated, enabling unsupervised identification of quantum phases.

\section{Result}

\begin{figure*}[htpb]
\includegraphics[width=1\linewidth]{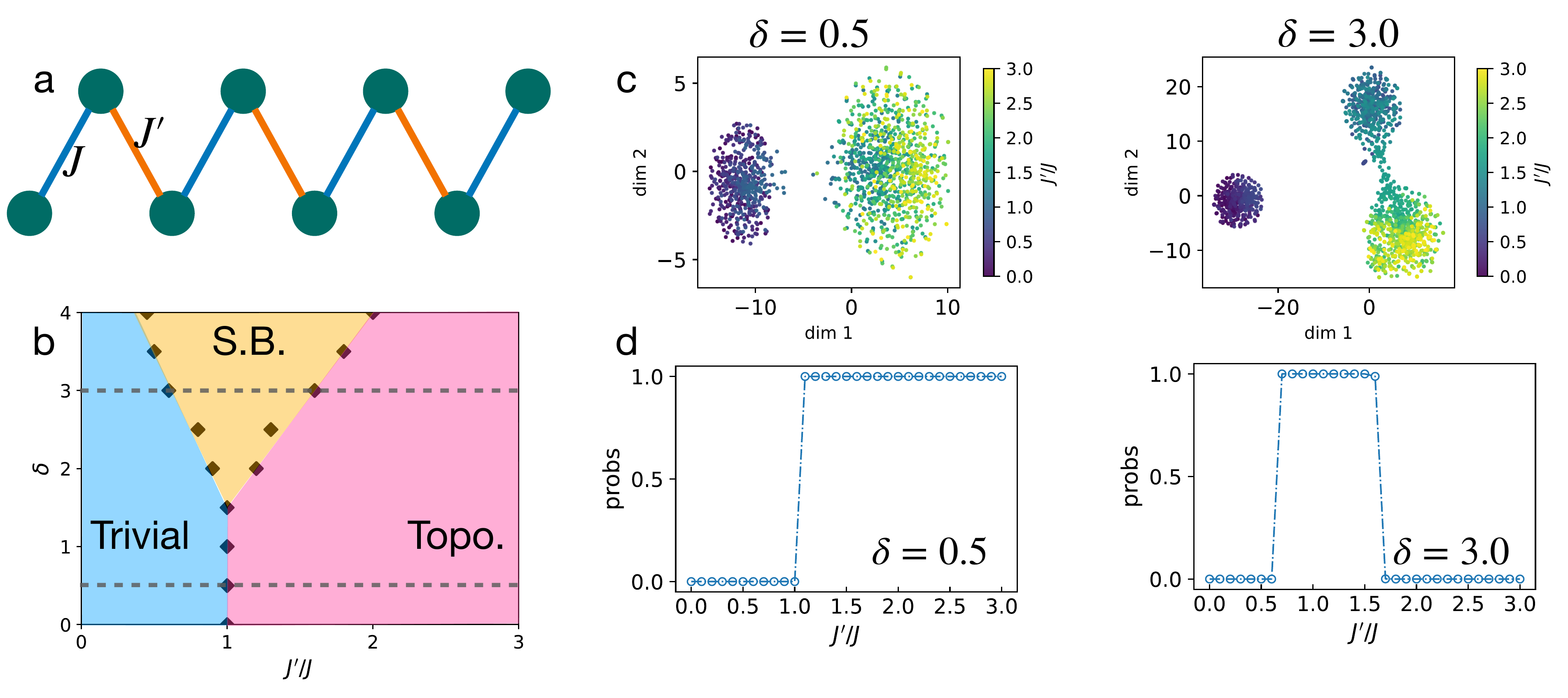}
\caption{ (a) Schematic representation of the extended SSH model Hamiltonian, illustrating the intra-cell hopping $J$, inter-cell hopping $J'$.  (b) Phase diagram where the theoretical phase boundaries are computed based on MBTI. Black diamond markers denote the phase transition points identified by our QRC-based unsupervised method. (c) Visualization of clustering in the t-SNE feature space, showing how the states naturally group according to their underlying phases. (d) Probabilities of phase assignment obtained from a Gaussian Mixture Model (GMM). To enhance clarity and readability, a sparse subset of the 1501 data points is plotted by selecting every 50th point.}
\label{fig:mr}
\end{figure*}

As the example of our framework, we study a parameterized strongly-correlated Hamiltonian system exhibiting quantum phase transitions. Specifically, we consider an $N=128$ sites open-boundary bond-alternating XXZ model, which can also be regarded as an interacting extended SSH model. [see Fig.~\ref{fig:mr}(a)]:

\begin{equation}
\begin{aligned}
H &= J\sum_{i \in \text{odd}} (X_i X_{i+1} + Y_i Y_{i+1} + \delta Z_i Z_{i+1}) \\
&+ J' \sum_{i \in \text{even}} (X_i X_{i+1} + Y_i Y_{i+1} + \delta Z_i Z_{i+1}),
\end{aligned}
\end{equation}

which hosts symmetry-protected topological (SPT) phases, symmetry-broken (S.B.) phases, and trivial phases. 

When $\delta = 0$, the system reduces to the well-know SSH model with a Bloch band structure. By tuning the parameters $J$, $J^\prime$, and $\delta$, the system can access different quantum phases. In conventional studies, one often characterizes the topological properties by computing or measuring the partial reflection MBTI~\cite{sciadvaaz3666}:
\begin{equation}
\tilde{\mathcal Z}_\mathcal R = 
\frac{\mathrm{Tr}(\rho_I \mathcal R_I)}{\sqrt{[\mathrm{Tr}(\rho_{I_1}^2) + \mathrm{Tr}(\rho_{I_2}^2)]/2}},
\end{equation}
where $\rho_I = \mathrm{Tr}_{S - I}(|\psi\rangle \langle \psi|)$ is the reduced density matrix of a sufficiently large non-local subsystem $I = I_1 \cup I_2$ ($S$ denotes all sites of the system, and each partition $I_1$ and $I_2$ contains $n$ sites). The non-local operator $\mathcal R_I$ spatially swaps the two partitions $I_1$ and $I_2$. 

For SPT phases, $\tilde{\mathcal Z}_\mathcal R = -1$; for trivial phases, $\tilde{\mathcal Z}_\mathcal R = 1$; and for symmetry-spontaneously-broken phases, $\tilde{\mathcal Z}_\mathcal R = 0$. This topological invariant requires reconstructing a sufficiently large non-local reduced density matrix and preparing a copy of the system for the swap operation, making it challenging both numerically and experimentally.

For simplicity, we set $J=1$, so that only two parameters, $\delta$ and $J^\prime$, need to be tuned. 
We scan the parameter space by choosing $\delta\in\{0.0,0.5,\dots,4.0\}$, and for each fixed $\delta$ we vary $J^\prime$ uniformly from $0$ to $3.0$ with a step size of $0.002$, yielding $1501$ parameter points. 
The corresponding ground states are obtained using the density matrix renormalization group (DMRG), and subsequently processed by our framework. 
Applying the our method, we obtain two-dimensional vector representations of the high-dimensional measurement data. 

To illustrate the results, we show the cases of $\delta = 0.5$ and $\delta = 3.0$ in Fig.~\ref{fig:mr}(c). 
For $\delta = 0.5$, the data cluster into two groups according to $J^\prime$, corresponding to the trivial and SPT phases. 
For $\delta = 3.0$, an additional SB phase emerges, and the data successfully cluster into three groups at the appropriate parameter regions. 
These clustering patterns are consistent with the predictions of the MBTI. 
We have verified that similar results hold for all other parameter values in the range $\delta \in [0.0, 4.0]$.

To precisely locate the phase transition points, we further apply Gaussian Mixture Model (GMM)~\cite{https://doi.org/10.1111/insr.12593} clustering to the two-dimensional t-SNE embeddings. 
Since GMM requires the number of clusters as an input, this information is obtained directly from the clustered structure revealed by our framework. 
After GMM fitting, we obtain for each data point both a discrete cluster label and the associated probability. 
At phase transition points, the assignment probability of data points switches abruptly between clusters (i.e., from $0$ to $1$). 
This provides a reliable way to determine the phase boundaries. 
In Fig.~\ref{fig:mr}d we show the GMM probability maps for $\delta = 0.5$ and $\delta = 3.0$, where the phase transition points can be clearly identified. 
Consistently, we obtain accurate transition points across the full parameter range $\delta \in [0.0, 4.0]$, and the extracted phase boundaries are marked in the phase diagram of Fig.~\ref{fig:mr}(b).

Regarding the circuit parameters, in the results shown in Fig.~\ref{fig:mr}, we set $g=0.96$, which drives the circuit evolution into the DTC regime, and we choose $D=25$ layers of Floquet evolution $U_F$. It is important to emphasize that not all parameter choices lead to meaningful representations.

\section{Neccessity of DTC Evolution}
We now discuss the necessity of using the DTC circuit. Previous theoretical studies have investigated MBL-DTC circuits from the perspective of information scrambling~\cite{PRXQuantum.2.030346}. For the specific circuit structure we employ, the out-of-time-order correlator (OTOC) between the first and last sites shows that, in the thermal regime, local information rapidly spreads across the entire systemz~\cite{Li2025023202}. In contrast, in the MBL-DTC regimes, local information exhibits almost no decay. From this perspective, one might naively consider the DTC circuit as an approximate ``identity'' or a very shallow thermal circuit. However, for our purpose, the DTC circuit plays a much more critical role.

\begin{figure}[htpb]
    \centering
    \includegraphics[width=1.0\linewidth]{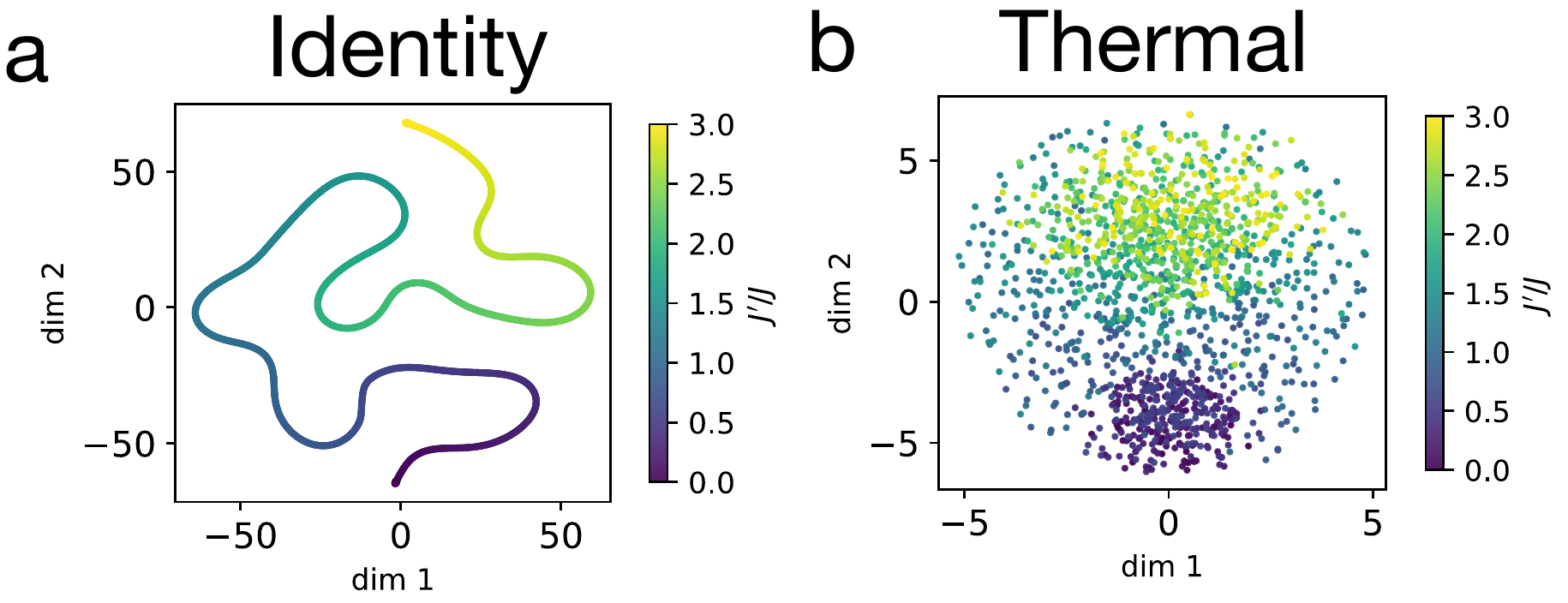}
    \caption{ Illustration of t-SNE in the non-MBL DTC region for $\delta = 3.0$. 
        (a) ``Identical data'' case, where t-SNE is applied directly to the ground-state measurements of $\langle Z_i \rangle$ and $\langle Z_i Z_{i+1} \rangle$. (b) t-SNE after evolution through the thermal region, highlighting how the feature space reorganizes under thermal dynamics. circuit parameters $g=0.5$, $D=5$.
   }
    \label{fig:ndtc}
\end{figure}

To illustrate this point, we examine the case of $\delta = 3.0$. If we perform no circuit evolution and directly measure local $\langle Z_i \rangle$ and $\langle Z_i Z_{i+1} \rangle$ on the ground states, which we refer to as the “identity data.” The resulting t-SNE embedding [Fig.~\ref{fig:ndtc}(a)] shows that all data points form a continuous curve in the reduced 2D space, with no clear clustering. This indicates that if one directly applies classical unsupervised learning to the ground states without any prior quantum processing, it is completely unable to resolve the quantum phase transitions. Next, if we evolve the states through a thermal circuit ($g=0.5$), even for an extremely shallow depth ($D=5$), the t-SNE embedding [Fig.~\ref{fig:ndtc}(b)] still fails to produce three distinct clusters; the data remain highly mixed, completely failing to reproduce the clustering observed in the DTC or MBL regimes.

A preliminary analysis suggests the following. t-SNE preserves local neighborhood structure: similar points cluster together, while dissimilar points are separated. For identity data [Fig.~\ref{fig:ndtc}(a)], features from different phases correspond to different segments on the same manifold, with insufficient local differences to form clear clusters. Under MBL-DTC evolution [Fig.~\ref{fig:ndtc}(b)], local neighborhood statistics between phases are enhanced, enabling t-SNE to capture phase separation more effectively. The MBL-DTC evolution can be interpreted as a nonlinear feature map that distorts and stretches the original manifold, increasing distinguishability of local structures in the embedding space. In contrast, the thermal regime introduces stronger distortions that, while nonlinear, can obscure distinctions when multiple phases need to be discriminated.

These observations confirm that either MBL or DTC dynamics are essential for the circuit. Moreover, since the DTC circuit has been shown to possess superior robustness against noise, it is the more suitable choice for implementation on realistic NISQ devices.

\section{Analysis}

Why does this behavior occur? To answer this question, we conducted further analysis. 

First, for ground states in different phases, local measurements $\langle Z_i \rangle$ and $\langle Z_i Z_{i+1} \rangle$ already contain some phase-related information. For instance, in the topological phase, the presence of edge states leads to $\langle Z_1 \rangle$ and $\langle Z_\text{end} \rangle$ being distinct from their values in the trivial phase, However, for the S.B. state, examining only $\langle Z_i \rangle$ can also reveal non-trivial features, and even $\langle Z_1 \rangle$ and $\langle Z_{\mathrm{end}} \rangle$ may take non-trivial values. Although these behaviors differ from those of the topological state, such distinctions are difficult for the machine to discern, which results in a continuous manifold in the reduced embedding space. Thus this alone is insufficient to fully separate the topological, symmetry-broken, and trivial phases. In a certain sense, this may also explain why, in earlier preliminary works, methods such as diffusion maps could directly learn the manifold structure of the Hamiltonian—for example, in the SSH model—and thereby extract phase transition–related information. The reason is that the manifold of the Hamiltonian directly reflects the nature of the ground state, and such models typically involve only the distinction between topological and trivial phases, making it much easier for the machine to identify their differences.

\begin{figure}[htpb]
    \includegraphics[width=1\linewidth]{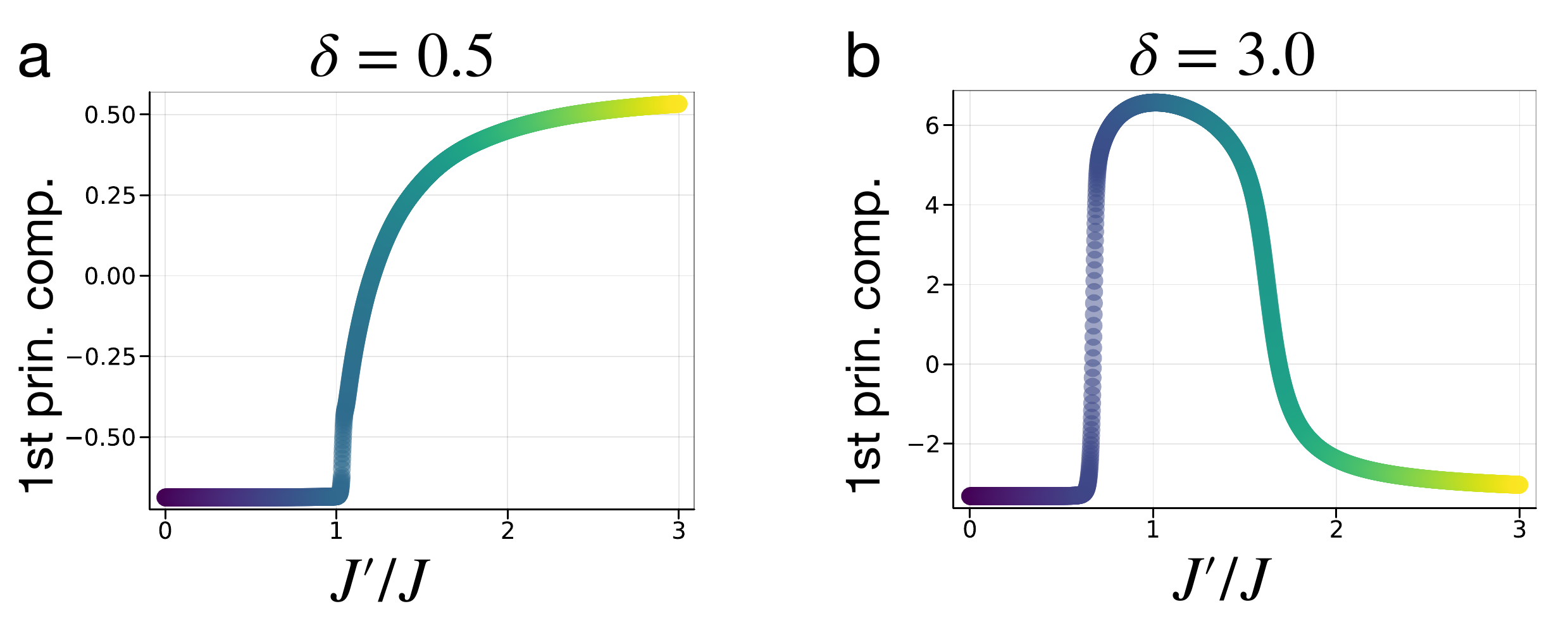}
    \caption{PCA of ground-state measurements of $\langle Z_i \rangle$ and $\langle Z_i Z_{i+1} \rangle$, showing the first principal component. 
        (a) $\delta = 0.5$. 
        (b) $\delta = 3.0$.}
    \label{fig:pca}
\end{figure}

We further support this viewpoint by performing principal component analysis (PCA)~\cite{Greenacre2022PCA} on the feature vectors constructed solely from the ground-state values of $\langle Z_i \rangle$. As shown in Fig.~\ref{fig:pca}. We find that, in the trivial phase, all feature vectors ``collapse'' to essentially the same value along one principal component. When the system hosts only trivial and topological phases, e.g., at $\delta = 0.5$ [Fig.~\ref{fig:pca}(a)], the separation between trivial and topological states is evident. However, at $\delta = 3.0$, where the system supports three distinct phases, although the trivial states still ``collapse'' along a single principal component, the feature vectors corresponding to the symmetry-broken and topological phases no longer exhibit such clear separation [Fig.~\ref{fig:pca}(b)].

\begin{figure*}
    \centering
    \includegraphics[width=1\linewidth]{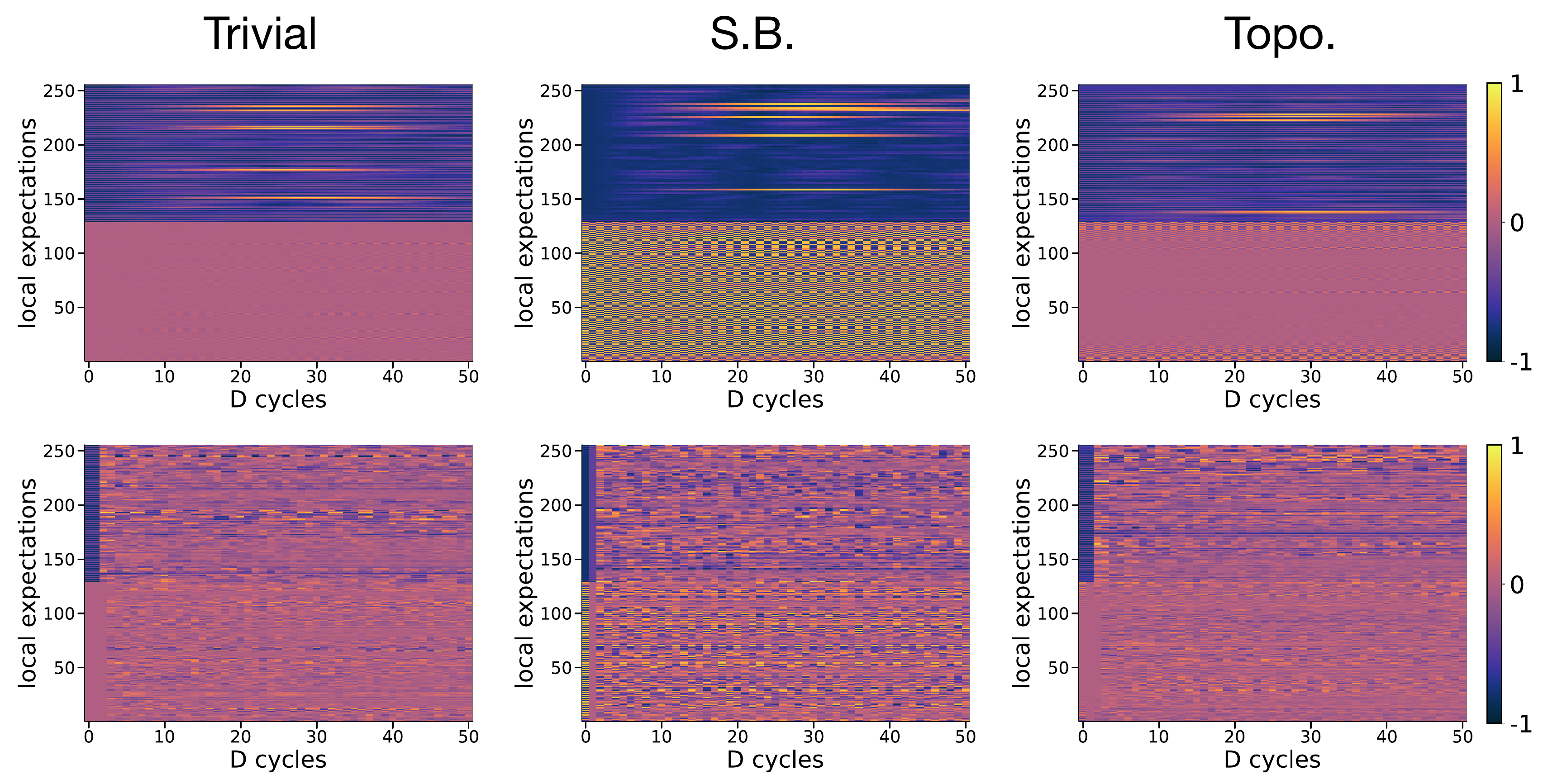}
    \caption{  Illustration of long-time evolution ($0$-$50$ cycles) for three representative states under different circuits. 
        The upper panels show DTC evolution ($g=0.96$) and the lower panels show thermal evolution ($g=0.5$). 
        From left to right, the states are: trivial ($J^\prime/J = 0.9$, $\delta = 1.0$), SB ($\delta = 2.0$, $J^\prime/J = 1.0$), and topological ($\delta = 1.0$, $J^\prime/J = 1.1$). 
        The vertical axis indices 1–128 correspond to $\langle Z_i \rangle$ ($i=1,\dots,128$), and 129–255 correspond to $\langle Z_i Z_{i+1} \rangle$ ($i=1,\dots,127$).
 }
    \label{fig:dynamicals}
\end{figure*}

Second, the evolution of ground states under the MBL-DTC circuit differs between phases. For example, $\langle Z_1 \rangle$ and $\langle Z_\text{end} \rangle$ exhibit period-doubled oscillations in the DTC regime, whereas they are rapidly erased in the thermal regime. Second, the evolution of ground states under the MBL-DTC circuit differs between phases. To illustrate this, we specifically selected three representative ground states that are close to each other in parameter space, each corresponding to a different phase: $\delta=1.0,  J^\prime/J=0.9$ for the trivial phase; $\delta=1.0,  J^\prime/J=1.1$ for the SPT phase; and $\delta=2.0,  J^\prime/J=1.0$ for the S.B. phase. We then studied their dynamics under different evolution regimes, as shown in Fig.~\ref{fig:dynamicals}. We perform DTC evolution with $g=0.96$ and thermal evolution with $g=0.5$. In Fig.~\ref{fig:dynamicals}, the vertical axis index from 1 to 128 corresponds to $\langle Z_i \rangle$, and indices 129 to 255 correspond to $\langle Z_i Z_{i+1} \rangle$ for $i=1$ to 127. Under DTC evolution, the initial local $Z$ information is largely preserved, with only limited erasure, whereas under thermal evolution, this information is completely erased within a very short time. Interestingly, the behavior of the $Z_iZ_{i+1}$ correlations across different regions exhibits significant variations and cannot be simply described in terms of erasure or preservation. This observation naturally leads to the next point of discussion.

It is also important to note that our high-dimensional feature vectors include not only $\langle Z_i \rangle$ but also $\langle Z_iZ_{i+1} \rangle$. The behavior of $\langle Z_iZ_{i+1} \rangle$ is significantly more complex than that of $\langle Z_i \rangle$ and cannot be fully characterized by the global scrambling from the first to the last site. A more detailed analysis of $\langle Z_iZ_{i+1} \rangle$ will be presented in Appendix~\ref{app:1}.

Taken these three mechanisms together, cause the measured data, after DTC evolution and t-SNE dimensionality reduction based on local neighborhood relationships in Eq.~\ref{equ:pca}, to form well-separated clusters corresponding to different phases in Fig.~\ref{fig:mr}. This enables us to successfully perform unsupervised learning and extract the phase transition information of the model.

\section{Discussion}

We have shown that SPT quantum phase transitions in the extend SSH Model can be efficiently identified through unsupervised learning based only on local $Z$-related expectation values measured from DTC circuits, which are known to be robust on NISQ devices. Compared with heuristic methods such as QCNNs that require parameter tuning, or tomography-based approaches relying on randomized POVM measurements, our scheme avoids both circuit optimization and expensive state reconstruction, thus offering a clear experimental advantage. Furthermore, the framework naturally generalizes to models with multi-site interactions e.g., the cluster-Ising model, where higher-order correlators such as $\langle Z_i Z_{i+1} Z_{i+2} \rangle$ can still be directly obtained from the same set of $n$-shot computational-basis measurements without extra overhead. This scalability highlights the practicality of our approach for probing quantum phase transitions on near-term quantum devices.

Our framework relies on the stability provided by many-body localization (MBL), and thus its applicability is presently limited to one-dimensional systems. Nevertheless, recent studies suggest that prethermal dynamics or certain topologically protected DTC modes can also be realized in two dimensions\cite{Jin2025Topological}, indicating a promising path toward higher-dimensional extensions. More importantly, our work introduces a new perspective by organically combining two seemingly distinct quantum phenomena—topological states and MBL-stabilized DTC dynamics—and exploring them through machine-learning techniques rather than relying solely on conventional physical quantities such as entropy. This interdisciplinary combination may not only lead to novel discoveries but also inspire new directions in the study of strongly correlated quantum many-body systems.

\section*{Code and Data Availability}
Source code for an efficient implementation of the proposed procedure is available at \texttt{https://github.com/zipeilee/qrc-phase} upon publication. Source data are available for this paper. All other data that support the plots within this paper and other
findings of this study are available from the corresponding author upon reasonable request.

\begin{acknowledgments}
We thank Shiwei Zhang for helpful discussions, the developers of \texttt{iTensors.jl}~\cite{itensor}, which we used for both DMRG and quantum circuit simulations, and OpenAI’s GPT-5 for assistance with manuscript writing and polishing. We also acknowledge the Mindquantum~\cite{xu2024mindspore} for supporting small-scale circuit simulations and measurements. This work was supported by the Beijing Institute of Technology Research Fund Program under Grant No. 2024CX01015.
\end{acknowledgments}

\appendix*
\section{Detailed Analysis of local expectations}\label{app:1}

In the MBL phase, each lattice site hosts a quasi-local conserved operator $\tau_i^z$ (an l-bit), which commutes exactly with the Hamiltonian and whose weight decays exponentially away from site $i$. In the physical Pauli-$Z$ basis, $\tau_i^z$ can be expanded as~\cite{Abanin_2019}
\begin{equation}
\tau_i^z = \sum_{S \subset \Lambda} \alpha_{i,S} \prod_{j \in S} Z_j,
\end{equation}
where the sum runs over all subsets $S$ of lattice sites $\Lambda$, and the coefficients $\alpha_{i,S}$ are quasi-local, satisfying $|\alpha_{i,S}| \lesssim C e^{-r(S)/\xi}$ with $r(S)$ the maximal distance from site $i$ to any site in $S$, $C$ a constant, and $\xi$ the localization length. In a conventional diagonal gauge, the expansion is dominated by short-range terms,
\begin{equation}
\tau_i^z \approx \sum_{j} \alpha^{(1)}_{i;j} Z_j 
+ \sum_{j<k} \alpha^{(2)}_{i;jk} Z_j Z_k 
+ \sum_{j<k<l} \alpha^{(3)}_{i;jkl} Z_j Z_k Z_l + \cdots,
\end{equation}
where $\alpha^{(1)}_{i;j}$ are the single-body coefficients, $\alpha^{(2)}_{i;jk}$ are the two-body coefficients, and $\alpha^{(3)}_{i;jkl}$ are the three-body or higher-order coefficients, which decay exponentially with the support radius of the operators.

Conversely, physical operators can also be expanded in the l-bit basis. Truncating to single- and two-body terms, the expansions of the single-site $Z_i$ and two-site $Z_i Z_{i+1}$ operators read
\begin{align}
Z_i &\approx \sum_j a^{(1)}_{i;j} \tau_j^z 
+ \sum_{j<k} a^{(2)}_{i;jk} \tau_j^z \tau_k^z + \cdots,
\end{align}
\begin{equation}
\begin{aligned}
Z_i Z_{i+1} &\approx \sum_{j<k} b^{(2)}_{i;jk} \tau_j^z \tau_k^z 
+ \sum_j b^{(1)}_{i;j} \tau_j^z 
+ \sum_{j<k<l} b^{(3)}_{i;jkl} \tau_j^z \tau_k^z \tau_l^z \\ &+ \cdots,
\end{aligned}
\end{equation}
where the leading coefficients $a^{(1)}_{i;i}$ and $b^{(2)}_{i;i,i+1}$ indicate that the physical operators primarily overlap with the conserved l-bit components, while the remaining coefficients represent multi-body dressing or non-conserved contributions.

From these expansions, one can see that the expectation value of a single-site operator $Z_i$ is mainly determined by the conserved component of a single l-bit, and therefore exhibits relatively stable oscillations or slow decay under MBL or DTC evolution. In contrast, the two-site operator $Z_i Z_{i+1}$ contains additional single- and three-body contributions, making its time evolution more sensitive to non-conserved components, which can manifest as spectral broadening or envelope decay. For higher-order operators such as $ZZZ$ or $ZZZZ$, the situation is similar: each additional site introduces more quasi-local l-bit contributions, making their temporal structure more complex and potentially further modifying their amplitudes. and their decay exponentially around site $i$.

Based on the exact l-bit construction in the previous work, these quasi-local operators $\tau_i^z$ exhibit weights that decay exponentially with distance, and their average localization length decreases as the disorder strength increases. Near the MBL-to-thermal transition, the distribution of localization lengths develops heavy tails. This indicates that, although the system as a whole remains localized, some operators can have a relatively large spatial support, thereby affecting the dynamics of multi-body expectation values such as $\langle Z_iZ_i+1\rangle$ and $\langle Z_iZ_{i+1}Z_{i+2}\rangle$. In summary, the differences between local operators like $Z_i$ and $Z_iZ_{i+1}$ can be naturally understood in terms of their expansion in the l-bit basis and the distribution of their weights.

Furthermore, in the context of DTC dynamics, the effective Hamiltonian exhibits a $\pi$-pairing of all eigenlevels~\cite{PRXQuantum.2.030346, Mi_2021}. This leads to subharmonic (double-period) oscillations in the expectation values of single-site operators $\tau_i^z$, or approximately $Z_i$, reflecting the DTC response. However, for two-site operators such as $Z_i Z_j$, the product involves two $\tau^z$ operators. Even when the primary overlap of each $\tau^z$ with the corresponding $Z$ is large, the double-period signal does not simply carry over; instead, the expectation value exhibits the fundamental period of the drive. This illustrates that while DTC signatures are naturally strong in single-site observables, multi-site correlators inherit a more complex time-dependence and do not generically exhibit the same subharmonic response.

Ultimately, both mechanisms are reflected in Fig.~\ref{fig:dynamicals}: during the DTC evolution, most $\langle Z_i Z_{i+1} \rangle$ exhibit only small periodic variations, but due to the more complex mixing of multi-body operator components, a few $\langle Z_i Z_{i+1} \rangle$ show pronounced changes, illustrating the combined effect of the l-bit structure and the dynamics of multi-body operators. This helps capture the characteristic features of different phases, facilitating the clustering of a batch of features from a continuous manifold when performing t-SNE.

Notably, Ref.~\cite{Kobayashi2025QRP} claims to study quantum phase transitions using ``quantum reservoir probing'' (QRP). Although inspired by QRC, their method essentially performs a quench of a trivial initial state under the target Hamiltonian and only borrows measurement observables inspired by QRC, without fundamentally departing from conventional quench protocols~\cite{Mitra_2018}. In contrast, our method is fundamentally different. The workflow: inputting quantum states, processing through the reservoir, performing local measurements, and applying post-processing—maintains independence among each stage, strictly adhering to QRC principles. We further innovate by introducing non trivial quantum data as input and replacing linear regression with unsupervised learning, thereby extending the scope of QRC. Moreover, implementing quench dynamics for strongly correlated many-body Hamiltonians on NISQ devices—such as via Trotterization—is operationally demanding and highly susceptible to noise. In contrast, our approach employs a hardware-efficient circuit and leverages the intrinsic noise robustness of discrete time crystals, making it far more suitable for practical implementation.

\bibliography{refs}

\end{document}